\documentclass[12pt,preprint,notoc,nohyper]{KSJCAP} 

\usepackage{epsfig,multicol,bbm}

\newcommand\fverb{\setbox\pippobox=\hbox\bgroup\verb}
\newcommand\fverbdo{\egroup\medskip\noindent%
            \fbox{\unhbox\pippobox}\ }
\newcommand\fverbit{\egroup\item[\fbox{\unhbox\pippobox}]}
\newbox\pippobox

\title{Charged fermions tunneling from accelerating and rotating black holes}

\author{Mudassar Rehman and K. Saifullah  \\

Department of Mathematics, Quaid-i-Azam University, Islamabad,
Pakistan \\

Electronic address: \email{saifullah@qau.edu.pk}}

\preprint{}  

\abstract{We study Hawking radiation of charged fermions from
accelerating and rotating black holes with electric and magnetic
charges. We calculate the tunneling probabilities of incoming and
outgoing fermionic particles and find the Hawking temperature of
these black holes. We also provide an explicit expression of the
classical action for the massive and massless particles in the
background of these black holes.}

\begin{document}

\section{Introduction}

Recently there has been a lot of interest in studying the Hawking
radiation \cite{Hawk1} as a phenomenon of quantum tunneling from
black holes \cite{PW, KW, Pa}. In one approach the tunneling rate of
particles coming out from the black hole horizon is calculated from
the imaginary part of their classical action \cite{SP, SPS}.
Following this approach radiation of scalar particles, and charged
and uncharged fermions was studied for the Kerr, Kerr-Newman,
Taub-NUT, G\"{o}del, dilatonic black holes and those with
acceleration and rotation \cite{KM06, KM08a, KM08b, ZL, LRW, CJZ,
Ji, ZY, DJ, YY, GS}.

In this paper we study tunneling of charged fermions from
accelerating and rotating black holes with electric and magnetic
charges \cite{PK, GP05, GP06, DL03a, DL03b, BS}. These black holes
have acceleration horizons also in addition to two rotation
horizons. We find the tunneling probabilities of these particles
using the WKB approximation and then calculate temperature at the
horizons. The paper is organized as follows. In Section 2 we
describe the spacetime for accelerating and rotating black holes
with electric and magnetic charges and their basic properties.
Section 3 deals with the solution of charged Dirac equation in the
background of these black holes and tunneling temperature is
calculated. In Section 4 we provide the exact form of the action for
massless as well as massive particles. We conclude briefly in
Section 5.

\section{Accelerating and rotating black holes with electric and magnetic charges}

The Pleba\'{n}ski and Demia\'{n}ski metric \cite{GP05, GP06, PD}
covers a large family of spacetimes which include the well-known
black hole solutions like Schwarzschild, Reissner-Nordstr\"{o}m,
Kerr, Kerr-Newman, Kerr-NUT and many others. This family includes,
in particular, solutions for accelerating and rotating black holes
also with electromagnetic field, which in spherical polar
coordinates can be written as

\begin{eqnarray}
ds^{2} &=&-\frac{1}{\Omega ^{2}}\left[ \frac{Q}{\rho
^{2}}dt^{2}+\frac{ Qa^{2}\sin ^{4}\theta }{\rho ^{2}}d\phi
^{2}-\frac{Q}{\rho ^{2}}2a\sin ^{2}\theta dtd\phi -\frac{\rho
^{2}}{Q}dr^{2}-\frac{\rho ^{2}}{P}d\theta
^{2}\right.  \label{1.2} \\
&&\left. -\frac{P\sin ^{2}\theta }{\rho ^{2}}a^{2}dt^{2}-\frac{P\sin
^{2}\theta }{\rho ^{2}}\left( r^{2}+a^{2}\right) ^{2}d\phi
^{2}+\frac{P\sin ^{2}\theta }{\rho ^{2}}2a\left( r^{2}+a^{2}\right)
dtd\phi \right] , \nonumber
\end{eqnarray}
where

\begin{eqnarray}
\Omega &=&1-\alpha r\cos \theta , \label{3.6} \\
\rho ^{2} &=&r^{2}+a^{2}\cos ^{2}\theta ,  \label{3.7} \\
P &=&1-2\alpha M\cos \theta +\left[ \alpha ^{2}\left(
a^{2}+e^{2}+g^{2}\right) \right] \cos ^{2}\theta ,  \label{3.8} \\
Q &=&\left[ \left( a^{2}+e^{2}+g^{2}\right) -2Mr+r^{2}\right] \left(
1-\alpha ^{2}r^{2}\right) .  \label{3.9}
\end{eqnarray}
Here $M$ is the mass of the black hole, $e$ and $g$ are its electric
and magnetic charges, $a$ is angular momentum per unit mass, and
$\alpha$ is the acceleration of the black hole. We write the metric
(\ref{1.2}) in a more convenient form as

\begin{eqnarray}
ds^{2} &=&-\left( \frac{Q-a^{2}P\sin ^{2}\theta }{\rho ^{2}\Omega
^{2}} \right) dt^{2}+\left( \frac{\rho ^{2}}{Q\Omega ^{2}}\right)
dr^{2}+\left(
\frac{\rho ^{2}}{P\Omega ^{2}}\right) d\theta ^{2}  \nonumber \\
&&+\left( \frac{\sin ^{2}\theta \left[ P\left( r^{2}+a^{2}\right)
^{2}-a^{2}Q\sin ^{2}\theta \right] }{\rho ^{2}\Omega ^{2}}\right)
d\phi ^{2}
\nonumber \\
&&-\left( \frac{2a\sin ^{2}\theta \left[ P\left( r^{2}+a^{2}\right)
-Q\right] }{\rho ^{2}\Omega ^{2}}\right) dtd\phi .  \label{2}
\end{eqnarray}
Now we use the notation of Ref. \cite{KM08a} and write the above
metric as
\begin{equation}
ds^{2}=-f\left( r,\theta \right) dt^{2}+\frac{dr^{2}}{g\left(
r,\theta \right) }+\Sigma \left( r,\theta \right) d\theta
^{2}+K\left( r,\theta \right) d\phi ^{2}-2H\left( r,\theta \right)
dtd\phi ,  \label{3}
\end{equation}
where the new functions are

\begin{eqnarray}
f\left( r,\theta \right) &=&\left( \frac{Q-a^{2}P\sin ^{2}\theta
}{\rho
^{2}\Omega ^{2}}\right) ,  \label{3.1} \\
g\left( r,\theta \right) &=&\frac{Q\Omega ^{2}}{\rho ^{2}},
\label{3.2} \\
\Sigma \left( r,\theta \right) &=&\left( \frac{\rho ^{2}}{P\Omega
^{2}}
\right) ,  \label{3.3} \\
K\left( r,\theta \right) &=&\left( \frac{\sin ^{2}\theta \left[
P\left( r^{2}+a^{2}\right) ^{2}-a^{2}Q\sin ^{2}\theta \right] }{\rho
^{2}\Omega ^{2}}
\right) ,  \label{3.4} \\
H\left( r,\theta \right) &=&\left( \frac{a\sin ^{2}\theta \left[
P\left( r^{2}+a^{2}\right) -Q\right] }{\rho ^{2}\Omega ^{2}}\right)
.  \label{3.5}
\end{eqnarray}
The electromagnetic vector potential for these black holes is given
by \cite{GP06}
\begin{equation}
A=\frac{-er\left[ dt-a\sin ^{2}\theta d\phi \right] -g\cos \theta
\left[ adt-\left( r^{2}+a^{2}\right) d\phi \right] }{r^{2}+a^{2}\cos
^{2}\theta }. \label{4}
\end{equation}
The event horizons of these black holes are obtained by putting
$g(r,\theta)=0$, which gives their location at

\begin{equation}
r=\frac{1}{\alpha \cos \theta },r=\pm \frac{1}{\alpha },\textrm{ and
} r_{\pm }=M\pm \sqrt{M^{2}-a^{2}-e^{2}-g^{2}}, \label{6}
\end{equation}
where $r_{\pm }$ represent the outer and inner horizons
corresponding to the Kerr-Newman black holes. The other horizons are
acceleration horizons. In this paper we study only the non-extremal
black holes i.e. those cases where the quantity under the radical
sign in Eq. (\ref{6}) remains positive. The angular velocity at the
black hole horizon \cite {KM08a, GS}
\begin{equation}
\Omega _{H}=\frac{H\left( r_{+},\theta \right) }{K\left(
r_{+},\theta \right) },  \label{6.1}
\end{equation}
becomes
\begin{equation}
\Omega _{H}=\frac{a}{\left( r_{+}^{2}+a^{2}\right) }.  \label{7}
\end{equation}
We define another function which will be needed later
\[
F\left( r,\theta \right) =f\left( r,\theta \right)
+\frac{H^{2}\left( r,\theta \right) }{K\left( r,\theta \right) }.
\]
Using values of $f$, $K$ and $H$ this takes the form
\begin{equation}
F\left( r,\theta \right) =\frac{PQ\rho ^{2}}{\Omega ^{2}\left[
P\left( r^{2}+a^{2}\right) ^{2}-a^{2}Q\sin ^{2}\theta \right] }.
\label{8}
\end{equation}

\section{Tunneling of charged fermions}

For studying tunneling of charged fermions, we will solve the Dirac
equation for accelerating and rotating black holes with electric and
magnetic charges. The covariant Dirac equation with electric charge
$q$ is

\begin{equation}
i\gamma ^{\mu }\left( D_{\mu }-\frac{iq}{\hbar }A_{\mu }\right) \Psi
+\frac{m }{\hbar }\Psi =0,  \label{9}
\end{equation}
where $m$ is the mass of the fermion particles and $D_{\mu
}=\partial _{\mu }+\Omega _{\mu }$, with $\Omega _{\mu
}=\frac{1}{2}i\Gamma _{\mu }^{\alpha \beta }\Sigma _{\alpha \beta
}$, and $\Sigma _{\alpha \beta }=\frac{1}{4}i\left[ \gamma ^{\alpha
},\gamma ^{\beta } \right]$ is antisymmetric. The quantities $\gamma
$'$s$ are defined as

\begin{eqnarray*}
\gamma ^{t} &=&\frac{1}{\sqrt{F\left( r,\theta \right) }}\gamma
^{0}, \gamma ^{r}=\sqrt{g\left( r,\theta \right) }\gamma ^{3},
\gamma ^{\theta }=\frac{1}{\sqrt{\Sigma \left( r,\theta \right) }}
\gamma ^{1},  \nonumber \\
\gamma ^{\phi } &=&\frac{1}{\sqrt{K\left( r,\theta \right) }}\left(
\gamma ^{2}+\frac{H\left( r,\theta \right) }{\sqrt{F\left( r,\theta
\right) K\left( r,\theta \right) }}\gamma ^{0}\right) , \label{13}
\end{eqnarray*}
where the matrices $\gamma ^{\mu }$ $\left( \mu =0,1,2,3\right) $
are
\begin{eqnarray*}
\gamma ^{0} &=&\left(
\begin{array}{cc}
0 & I \\
-I & 0
\end{array}
\right) , \gamma ^{1}=\left(
\begin{array}{cc}
0 & \sigma ^{1} \\
\sigma ^{1} & 0
\end{array}
\right) ,  \nonumber \\
\gamma ^{2} &=&\left(
\begin{array}{cc}
0 & \sigma ^{2} \\
\sigma ^{2} & 0
\end{array}
\right) , \gamma ^{3}=\left(
\begin{array}{cc}
0 & \sigma ^{3} \\
\sigma ^{3} & 0
\end{array}
\right) .  \label{14}
\end{eqnarray*}
The Pauli matrices $\sigma ^{i}$ $\left( i=1,2,3\right) $ are
\begin{eqnarray*}\nonumber
\sigma ^{1}=\left(
\begin{array}{cc}
0 & 1 \\
1 & 0
\end{array}
\right)  \sigma ^{2}=\left(
\begin{array}{cc}
0 & -i \\
i & 0
\end{array}
\right) ,\sigma ^{3}=\left(
\begin{array}{cc}
1 & 0 \\
0 & -1
\end{array}
\right) .  \label{15}
\end{eqnarray*}
Note that

\begin{equation}
\gamma ^{5}=i\gamma ^{t}\gamma ^{r}\gamma ^{\theta }\gamma ^{\phi
}=\sqrt{ \frac{g}{FK\Sigma }}\left(
\begin{array}{cc}
-I+\frac{H}{\sqrt{FK}}\sigma ^{2} & 0 \\
0 & I+\frac{H}{\sqrt{FK}}\sigma ^{2}
\end{array}
\right) ,  \label{16}
\end{equation}
is the resulting $\gamma ^{5}$ matrix. The eigenvector of $\sigma
^{3}$ for the spin up and spin down cases are $\xi _{\uparrow
}=\left(
\begin{array}{c}
1 \\
0
\end{array} \right)$ and $\xi _{\downarrow }=\left(
\begin{array}{c}
0 \\
1
\end{array}
\right)$.
We take the following assumptions for the solution of the
spin up and spin down particles respectively

\begin{equation}
\Psi _{\uparrow }\left( t,r,\theta ,\phi \right) =\left(
\begin{array}{c}
A\left( t,r,\theta ,\phi \right) \xi _{\uparrow } \\
B\left( t,r,\theta ,\phi \right) \xi _{\uparrow }
\end{array}
\right) \exp \left[ \frac{i}{\hbar }I_{\uparrow }\left( t,r,\theta
,\phi \right) \right] ,  \label{17}
\end{equation}

\begin{equation}
\Psi _{\downarrow }\left( t,r,\theta ,\phi \right) =\left(
\begin{array}{c}
A\left( t,r,\theta ,\phi \right) \xi _{\downarrow } \\
B\left( t,r,\theta ,\phi \right) \xi _{\downarrow }
\end{array}
\right) \exp \left[ \frac{i}{\hbar }I_{\downarrow }\left( t,r,\theta
,\phi \right) \right] ,  \label{18}
\end{equation}
where $I_{\uparrow /\downarrow }$ denote the action of the emitted
spin up and spin down particles, respectively. We will only give
calculations for the spin up case; those for the spin down case are
similar, apart from the change in sign. Now using the antisymmetric
property i.e. $\left[ \gamma ^{\alpha },\gamma ^{\beta }\right] =0$,
if $\alpha =\beta $ and $\left[ \gamma ^{\alpha },\gamma ^{\beta
}\right] =-\left[ \gamma ^{\beta },\gamma ^{\alpha }\right] $, if
$\alpha \neq \beta $, we find that the Dirac Eq. (\ref{9}) reduces
to
\begin{equation}
i\gamma ^{\mu }\left( \partial _{\mu }-\frac{iq}{\hbar }A_{\mu
}\right) \Psi +\frac{m}{\hbar }\Psi =0.  \label{19}
\end{equation}
Substituting the ansatz (\ref{17}) and (\ref{18}) into the Dirac
equation we obtain after some algebra the following four equations

\bigskip
\begin{eqnarray}
0 &=&-B\left[ \frac{1}{\sqrt{F\left( r,\theta \right) }}\partial
_{t}I_{\uparrow }+\sqrt{g\left( r,\theta \right) }\partial
_{r}I_{\uparrow }+ \frac{H\left( r,\theta \right) }{K\left( r,\theta
\right) \sqrt{F\left(
r,\theta \right) }}\partial _{\phi }I_{\uparrow }\right.  \nonumber \\
&&\left. -\frac{1}{\sqrt{F\left( r,\theta \right)
}}qA_{t}-\frac{H\left( r,\theta \right) }{K\left( r,\theta \right)
\sqrt{F\left( r,\theta \right) }} qA_{\phi }\right] +mA ,  \label{28}
\end{eqnarray}

\bigskip
\begin{equation}
0=-B\left[ \frac{1}{\sqrt{\Sigma \left( r,\theta \right) }}\partial
_{\theta }I_{\uparrow }+\frac{i}{\sqrt{K\left( r,\theta \right)
}}\partial _{\phi }I_{\uparrow }-\frac{i}{\sqrt{K\left( r,\theta
\right) }}qA_{\phi }\right],  \label{29}
\end{equation}

\bigskip
\begin{equation}
\begin{array}{c}
0=A\left[ \frac{1}{\sqrt{F\left( r,\theta \right) }}\partial
_{t}I_{\uparrow }-\sqrt{g\left( r,\theta \right) }\partial
_{r}I_{\uparrow }+\frac{H\left( r,\theta \right) }{K\left( r,\theta
\right) \sqrt{F\left( r,\theta \right) }}
\partial _{\phi }I_{\uparrow }\right. \\
\left. -\frac{1}{\sqrt{F\left( r,\theta \right)
}}qA_{t}-\frac{H\left( r,\theta \right) }{K\left( r,\theta \right)
\sqrt{F\left( r,\theta \right) }} qA_{\phi }\right] +mB,
\end{array}
\label{30}
\end{equation}

\bigskip
\begin{equation}
0=-A\left[ \frac{1}{\sqrt{\Sigma \left( r,\theta \right) }}\partial
_{\theta }I_{\uparrow }+\frac{i}{\sqrt{K\left( r,\theta \right)
}}\partial _{\phi }I_{\uparrow }-\frac{i}{\sqrt{K\left( r,\theta
\right) }}qA_{\phi }\right].  \label{31}
\end{equation}
Now the metric coefficients do not depend on $t$ and $\phi $
coordinates therefore we can apply the standard ansatz \cite{KM08a,
KM08b, GS}

\begin{equation}
I_{\uparrow }=-Et+J\phi +W\left( r,\theta \right) ,  \label{32}
\end{equation}
where $E$ and $J$ denote the energy and angular momentum of the
emitted particle. Now, expanding $g\left( r,\theta \right) $ and
$F\left( r,\theta \right)$ using Taylor's theorem near the outer
horizon and neglecting squares and higher powers, as has been done
in the case of uncharged particles and neutral black holes
\cite{GS}, and substituting the values of $A_{t}(r_{+},\theta)$,
$A_{\phi}( r_{+},\theta)$, $\Sigma (r_{+},\theta)$ and
$K(r_{+},\theta)$ the above set of equations take the form

\begin{equation}
-B\left[ \frac{\left( -E+\Omega _{H}J+\frac{qer_{+}}{\left(
r_{+}^{2}+a^{2}\right) }\right) }{\sqrt{\left( r-r_{+}\right)
\partial _{r}F\left( r_{+},\theta \right) }}+\sqrt{\left(
r-r_{+}\right) \partial _{r}g\left( r_{+},\theta \right) }\left(
\partial _{r}W\right) \right] +mA=0,  \label{39}
\end{equation}

\begin{eqnarray}\nonumber
-B\left[ \sqrt{\frac{P\Omega ^{2}\left( r_{+},\theta \right) }{\rho
^{2}\left( r_{+},\theta \right) }}\partial _{\theta }W  \right. \hspace{85mm} \\
\left. +\frac{i\rho \left( r_{+},\theta \right) \Omega \left(
r_{+},\theta \right) }{\sqrt{\sin ^{2}\theta P\left(
r_{+}^{2}+a^{2}\right) ^{2}}}\left( J-q\left( \frac{ aer_{+}\sin
^{2}\theta +g\left( r_{+}^{2}+a^{2}\right) \cos \theta }{
r_{+}^{2}+a^{2}\cos ^{2}\theta }\right) \right) \right] =0,
\label{40}
\end{eqnarray}

\begin{equation}
A\left[ \frac{\left( -E+\Omega _{H}J+\frac{qer_{+}}{\left(
r_{+}^{2}+a^{2}\right) }\right) }{\sqrt{\left( r-r_{+}\right)
\partial _{r}F\left( r_{+},\theta \right) }}-\sqrt{\left(
r-r_{+}\right) \partial _{r}g\left( r_{+},\theta \right) }\partial
_{r}W\right] +Bm=0.  \label{41}
\end{equation}

\begin{eqnarray}\nonumber
-A\left[ \sqrt{\frac{P\Omega ^{2}\left( r_{+},\theta \right) }{\rho
^{2}\left( r_{+},\theta \right) }}\partial _{\theta }W \right.
\hspace{85mm}  \\
\left. +\frac{i\rho \left( r_{+},\theta \right)
\Omega \left( r_{+},\theta \right) }{\sqrt{\sin ^{2}\theta P\left(
r_{+}^{2}+a^{2}\right) ^{2}}}\left( J-q\left( \frac{ aer_{+}\sin
^{2}\theta +g\left( r_{+}^{2}+a^{2}\right) \cos \theta }{
r_{+}^{2}+a^{2}\cos ^{2}\theta }\right) \right) \right] =0.
\label{42}
\end{eqnarray}
It is possible to sperate $W(r,\theta)$ near the black hole horizon
as

\begin{equation}
W\left( r,\theta \right) =R\left( r\right) +\Theta \left( \theta
\right) . \label{43}
\end{equation}
First we solve Eqs. (\ref{39}) - (\ref {42}) for the massless case
i.e. $m=0$. Using the above separation and taking $m=0$ we get
\[
-B\left[ \frac{\left( -E+\Omega _{H}J+\frac{qer_{+}}{\left(
r_{+}^{2}+a^{2}\right) }\right) }{\sqrt{\left( r-r_{+}\right)
\partial _{r}F\left( r_{+},\theta \right) }}+\sqrt{\left(
r-r_{+}\right) \partial _{r}g\left( r_{+},\theta \right) }R^{\prime}
\left( r\right) \right] =0,
\]
which implies
\begin{equation}
R^{\prime} \left( r\right) =R^{\prime}_{+} \left( r\right)
=\frac{\left( E-\Omega _{H}J-\frac{qer_{+}}{\left(
r_{+}^{2}+a^{2}\right) }\right) \left( r_{+}^{2}+a^{2}\right)
}{2\left( r-r_{+}\right) \left( r_{+}-M\right) \left( 1-\alpha
^{2}r_{+}^{2}\right) } , \label{44}
\end{equation}
where $R_{+}$ corresponds to the outgoing solution. Similarly the
incoming solution is obtained as
\begin{equation}
R^{\prime}\left( r\right) =R^{\prime}_{-} \left( r\right)
=-\frac{\left( E-\Omega _{H}J-\frac{qer_{+}}{\left(
r_{+}^{2}+a^{2}\right) }\right) \left( r_{+}^{2}+a^{2}\right)
}{2\left( r-r_{+}\right) \left( r_{+}-M\right) \left( 1-\alpha
^{2}r_{+}^{2}\right) } . \label{45}
\end{equation}

The imaginary part of $R_{+}$ is
\begin{equation}
Im R_{+}=\frac{\pi }{2}\frac{\left( E-\Omega _{H}J-\frac{qer_{+}}{
\left( r_{+}^{2}+a^{2}\right) }\right) \left( r_{+}^{2}+a^{2}\right)
}{ \left( r_{+}-M\right) \left( 1-\alpha ^{2}r_{+}^{2}\right) }.
\label{46}
\end{equation}
Similarly

\begin{equation}
Im R_{-}=\frac{- \pi \left( E-\Omega _{H}J-\frac{qer_{+}}{ \left(
r_{+}^{2}+a^{2}\right) }\right) \left( r_{+}^{2}+a^{2}\right) }{
 2 \left( r_{+}-M\right) \left( 1-\alpha ^{2}r_{+}^{2}\right) } ,
\label{46.1}
\end{equation}
showing that
\begin{equation}
Im R_{+}=- Im R_{-}.  \label{46.2}
\end{equation}
The probabilities of crossing the horizon in each direction are
proportional to \cite{SP, SPS}
\begin{equation}
P_{emission} \propto \exp \left[ -2 Im I\right] =\exp \left[
-2\left( Im R_{+}+Im \Theta \right) \right] , \label{47}
\end{equation}

\begin{equation}
P_{absorption} \propto \exp \left[ -2Im I\right] =\exp \left[
-2\left( Im R_{-}+Im \Theta \right) \right] . \label{48}
\end{equation}
Thus the probability of a particle tunneling from inside to outside
the horizon is
\[
\Gamma \propto \frac{P_{emission}}{P_{absorption}}=\frac{ \exp
\left[ -2\left( Im R_{+}+Im \Theta \right) \right] }{\exp \left[
-2\left( Im R_{-}+Im \Theta \right) \right] },
\]
or
\[
\Gamma =\exp \left[ -2\left( Im R_{+}-Im R_{-}\right) \right] ,
\]
and using Eq. (\ref{46.2})
\[
\Gamma =\exp \left[ -4Im R_{+}\right] .
\]
which in view of Eq. (\ref{46}) becomes

\begin{equation}
\Gamma =\exp \left[ \frac{-2 \pi \left( r_{+}^{2}+a^{2}\right)
}{\left( r_{+}-M\right) \left( 1-\alpha ^{2}r_{+}^{2}\right) }\left(
E-\Omega _{H}J- \frac{qer_{+}}{\left( r_{+}^{2}+a^{2}\right)
}\right) \right] . \label{49}
\end{equation}
Comparing this with $\Gamma=\exp [-\beta E]$ where $\beta = 1/T_H$
we find that the Hawking temperature \cite{SP, SPS} is given by

\[
T_{H}=\frac{\left( r_{+}-M\right) \left( 1-\alpha
^{2}r_{+}^{2}\right) }{2\pi \left( r_{+}^{2}+a^{2}\right)} .
\]

Note that if we put acceleration equal to zero in formulae
(\ref{49}), we recover the tunneling probability for the Kerr-Newman
black hole \cite{KM06}. Similarly, putting rotation equal to zero
will yield the formula for the Reissner-Nordstr\"{o}m black hole. It
is worth mentioning here that from Eqs. (\ref{47})-(\ref{49}) it
seems that for some value of $E$ and $J$ it is possible that the
probabilities become greater than 1 and hence violate unitarity.
However, this does not happen because apart from the spatial
contribution the temporal part also contributes to the imaginary
part $Im(E\Delta t)$ of the action \cite{APGS, APS}. The time is
shifted by some imaginary amount which contributes both to
$P_{emission}$ and $P_{absorption}$ and results in a correct value
of $\Gamma$. If this contribution is not taken into account, one
gets the Hawking temperature twice as large as the original value,
the so-called \emph{factor of 2 issue} \cite{AAS, Pi}. For the
massive case ($m\neq 0$) we again obtain the same temperature which
is not unexpected \cite{GS}, as the behaviour of massive particles
near the black hole horizon is same as that of the massless.

\section{Calculation of the action}

In order to obtain the explicit expression for the action
$I_{\uparrow}$ in the spin-up case, we solve Eqs.
(\ref{39})-(\ref{42}) near the black hole horizon. On using Eq.
(\ref{43}) we can write Eq. (\ref{39}) in the form
\[
R^{\prime} \left( r\right) =\frac{mA}{B\sqrt{\partial _{r}g\left(
r_{+},\theta \right) \left( r-r_{+}\right) }}-\frac{\left( -E+\Omega
_{H}J+\frac{qer_{+}}{\left( r_{+}^{2}+a^{2}\right) }\right)
}{\sqrt{\partial _{r}F\left( r_{+},\theta \right) \partial
_{r}g\left( r_{+},\theta \right) }\left( r-r_{+}\right) }.
\]
Integrating with respect to $r$ we get
\begin{equation}
R\left( r\right) =R_{+}\left( r\right) =\int
\frac{mA}{B\sqrt{\partial _{r}g\left( r_{+},\theta \right) \left(
r-r_{+}\right) }}dr-\frac{\left( -E+\Omega
_{H}J+\frac{qer_{+}}{\left( r_{+}^{2}+a^{2}\right) }\right) }{
\sqrt{\partial _{r}F\left( r_{+},\theta \right) \partial _{r}g\left(
r_{+},\theta \right) }}\ln \left( r-r_{+}\right) .  \label{65}
\end{equation}
Similarly for the incoming particles we obtain from Eq. (\ref{41})
\begin{equation}
R\left( r\right) =R_{-}\left( r\right) =\int
\frac{mB}{A\sqrt{\partial _{r}g\left( r_{+},\theta \right) \left(
r-r_{+}\right) }}dr+\frac{\left( -E+\Omega
_{H}J+\frac{qer_{+}}{\left( r_{+}^{2}+a^{2}\right) }\right) }{
\sqrt{\partial _{r}F\left( r_{+},\theta \right) \partial _{r}g\left(
r_{+},\theta \right) }}\ln \left( r-r_{+}\right) .  \label{66}
\end{equation}
From Eq. (\ref{40}) (or (\ref {42})) on using Eq. (\ref{43}) we
obtain

\begin{eqnarray}\nonumber
\sqrt{\frac{P\Omega ^{2}\left( r_{+},\theta \right) }{\rho
^{2}\left( r_{+},\theta \right) }} \partial _{\theta }\Theta \hspace{95mm} \\
+\frac{i\rho \left( r_{+},\theta \right) \Omega \left( r_{+},\theta
\right) }{\sqrt{\sin ^{2}\theta \left[ P\left(
r_{+}^{2}+a^{2}\right) ^{2}\right] }}\left( J-q\left(
\frac{aer_{+}\sin ^{2}\theta +g\left( r_{+}^{2}+a^{2}\right) \cos
\theta }{r_{+}^{2}+a^{2}\cos ^{2}\theta }\right) \right) =0.
\label{67}
\end{eqnarray}
Substituting the values of $\rho$ and $P$ from Eqs. (\ref{3.7}) and
(\ref{3.8}) and after simplifying this yields
\begin{eqnarray*}
\partial _{\theta }\Theta &=&\frac{}{}\frac{
(iqaer_{+}+iJa^{2})\sin \theta }{(r_{+}^{2}+a^{2})\left[1-2\alpha
M\cos \theta +\alpha ^{2}\left(
a^{2}+e^{2}+g^{2}\right) \cos ^{2}\theta \right]} \\
&&+\frac{iqg\cos \theta -iJ}{\sin \theta \left[ 1-2\alpha M\cos
\theta +\alpha ^{2}\left( a^{2}+e^{2}+g^{2}\right) \cos ^{2}\theta
\right] }.
\end{eqnarray*}
Integrating with respect to $\theta $ we get
\begin{equation}
\Theta =\frac{iqaer_{+}+iJa^{2}}{r_{+}^{2}+a^{2}}I_{1}+I_{2},
\label{68}
\end{equation}
where $I_{1}$ and $I_{2}$ are obvious substitutions, which after
evaluating the integrals come out to be

\begin{equation}
I_{1}=-\frac{1}{2\alpha \sqrt{M^{2}-a^{2}-e^{2}-g^{2}}}\ln \left(
\frac{ \alpha \cos \theta \left( a^{2}+e^{2}+g^{2}\right) -M-\sqrt{
M^{2}-a^{2}-e^{2}-g^{2}}}{\alpha \cos \theta \left(
a^{2}+e^{2}+g^{2}\right) -M+\sqrt{M^{2}-a^{2}-e^{2}-g^{2}}}\right),
\label{68.1}
\end{equation}

\begin{eqnarray}
I_{2} &=&L_1 \ln \left( 1-\cos \theta \right) + L_2 \ln \left( 1+\cos \theta \right)  \nonumber \\
&&-L_3 \ln \left[ 1-2\alpha M\cos \theta +\alpha ^{2}\left(
a^{2}+e^{2}+g^{2}\right) \cos ^{2}\theta \right]  \nonumber \\
&&- L_4 \ln \left[ \frac{\alpha \left( a^{2}+e^{2}+g^{2}\right) \cos
\theta -M-\sqrt{M^{2}-a^{2}-e^{2}-g^{2}}}{\alpha \left(
a^{2}+e^{2}+g^{2}\right) \cos \theta
-M+\sqrt{M^{2}-a^{2}-e^{2}-g^{2}}}\right] , \label{69}
\end{eqnarray}
where $L_i$ are given below
\begin{eqnarray}
L_1 &=&\frac{iqg-iJ}{2\left[ 1-2\alpha M+\alpha ^{2}\left(
a^{2}+e^{2}+g^{2}\right) \right] },  \label{68.21} \\
L_2 &=&\frac{iqg+iJ}{2\left[ 1+2\alpha M+\alpha ^{2}\left(
a^{2}+e^{2}+g^{2}\right) \right] },  \label{68.22} \\
L_3 &=&\frac{ iqg\left( 1+\alpha ^{2}\left( a^{2}+e^{2}+g^{2}\right)
\right) -2\alpha MiJ}{2 \left[ 1-2\alpha M+\alpha ^{2}\left(
a^{2}+e^{2}+g^{2}\right) \right] \left[ 1+2\alpha M+\alpha
^{2}\left( a^{2}+e^{2}+g^{2}\right) \right] }, \label{68.23} \\
\nonumber L_4 &=&\frac{\left( a^{2}+e^{2}+g^{2}\right) \left[
-iJ\alpha ^{2}-iJ\alpha ^{4}\left( a^{2}+e^{2}+g^{2}\right)
+iqg\alpha ^{3}M\right] +2iJ\alpha ^{2}M^{2}-iqg\alpha M}{2\alpha
\sqrt{M^{2}-a^{2}-e^{2}-g^{2}}\left[ \left\{ 1+\alpha ^{2}\left(
a^{2}+e^{2}+g^{2}\right) \right\} ^{2}-4\alpha ^{2}M^{2} \right] } .
\\ \label{68.24}
\end{eqnarray}
Using these values of $I_1$ and $I_2$ in Eq. (\ref{68}) we get
$W(r,\theta)$ from Eq. (\ref{43}) and an explicit expression for the
action of the outgoing massive particles. Putting $m=0$ will give
action for the massless case. Similarly, we obtain results for the
incoming fermions as well.

\section{The acceleration horizon}

As mentioned earlier, apart from the rotation horizons, these black
holes admit acceleration horizons, $r_{\alpha }= 1/\alpha \cos
\theta$ and $\pm 1/\alpha$, as well. Here we calculate the tunneling
probability of fermions from these horizons. Proceeding as before we
evaluate angular velocity and radial derivatives of functions $g$
and $F$ at the acceleration horizons and obtain a corresponding set
of four equations like Eqs. (\ref{39}) to (\ref{42}). An analysis
similar to the one done earlier for the massive case yields for the
outgoing and incoming fermions

\begin{equation}
R_{+}=\frac{\pi i\left( E-\Omega _{\alpha }J- qer_{\alpha }/\left(
r_{\alpha }^{2}+a^{2}\right) \right) \left( r_{\alpha
}^{2}+a^{2}\right) }{2\alpha^2 r_{\alpha} \left[ r_{\alpha
}^{2}-2Mr_{\alpha }+\left( a^{2}+e^{2}+g^{2}\right) \right] },
\label{88}
\end{equation}

\begin{equation}
R_{-}=-\frac{\pi i\left( E-\Omega _{\alpha }J- qer_{\alpha }/\left(
r_{\alpha }^{2}+a^{2}\right) \right) \left( r_{\alpha
}^{2}+a^{2}\right) }{ 2\alpha^2 r_{\alpha} \left[ r_{\alpha
}^{2}-2Mr_{\alpha }+\left( a^{2}+e^{2}+g^{2}\right) \right] }.
\label{89}
\end{equation}
Thus the probability of tunneling comes out to be
\begin{equation}
\Gamma =\exp \left[ \frac{-2\pi \left( r_{\alpha }^{2}+a^{2}\right)
\left( E-\Omega _{\alpha }J- qer_{\alpha }/\left( r_{\alpha
}^{2}+a^{2}\right) \right) }{\alpha^2 r_{\alpha} \left[ r_{\alpha
}^{2}-2Mr_{\alpha }+\left( a^{2}+e^{2}+g^{2}\right) \right] }\right]
.  \label{90}
\end{equation}
As before we compare this expression with $\Gamma=\exp [-\beta E]$
to write the Hawking temperature at the acceleration horizon as

\[
T_{H}=\frac{\alpha^2 r_{\alpha} \left[ r_{\alpha }^{2}-2Mr_{\alpha
}+\left( a^{2}+e^{2}+g^{2}\right) \right] }{2\pi \left(
r_{\alpha}^{2}+a^{2}\right)} .
\]

\section{Conclusion}

Hawking radiation has been studied as a phenomenon of quantum
tunneling from black hole horizons using different methods including
the Newman-Penrose formalism and Hamilton-Jacobi method. We have
used the semi-classical approach to study tunneling of charged
fermions from accelerating and rotating black holes having electric
and magnetic charges. From the tunneling probability of the incoming
and outgoing particles we have worked out the Hawking temperature of
these black holes. We know that classically only the outgoing
particles face the barrier but in the semi-classical approach we
have followed both the incoming as well as outgoing particles
\emph{see} the horizon as the tunneling barrier.

Secondly, we note that the tunneling probability of charged fermions
from inside to outside the horizon does not depend upon the mass of
the fermion but its charge. The Hawking temperature depends upon
acceleration as well apart from the mass, rotation parameter and the
electric and magnetic charges of the black hole. However, the amount
of acceleration is restricted by the value of other parameters
\cite{DL03a, DL03b, BS}.

\acknowledgments

Some helpful comments from Douglas Singleton are gratefully
acknowledged.

\end{document}